\documentclass[letterpaper]{emulateapj}
\usepackage{apjfonts}
\usepackage{lscape} 
\usepackage{amsmath,amsfonts}
\usepackage{graphicx}
\usepackage[dvips]{color}
\usepackage[breaklinks=true]{hyperref}
\usepackage{natbib} 
\usepackage[all]{draftcopy}
\bibpunct[,]{(}{)}{;}{a}{}{,} 

\newcommand{\Npart}{\ensuremath{{\cal N}_{\star}}}

\newcommand{\msol}{\ensuremath{{\rm M}_{\odot}}}
\newcommand{\Msun}{\msol}
\newcommand{\Mstar}{\ensuremath{m_{\ast}}}

\newcommand{\pc}{\ensuremath{\mathrm{pc}}}

\newcommand{\kms}{\ensuremath{\mathrm{km\,s}^{-1}}}

\newcommand{\NB}{{\sc NBODY4}}
\newcommand{\LISA}{\it LISA}
\newcommand{\BBO}{\it BBO}
\newcommand{\tento}[2]{\ensuremath{{#1}\times 10^{#2}}}

\newcommand{\Rem}[1]{}

\shorttitle{IMBHs in colliding clusters: Implications for lower-frequency 
GW astronomy}
\shortauthors{Amaro-Seoane \& Freitag}

\begin{document}


\title{Intermediate-mass black holes in colliding clusters:\\ Implications
for lower-frequency gravitational-wave astronomy}


\author{Pau Amaro-Seoane\altaffilmark{1}\thanks{e-mail: Pau.Amaro-Seoane@aei.mpg.de},
Marc Freitag\altaffilmark{2}\thanks{e-mail: freitag@ast.cam.ac.uk}}
\altaffiltext{1}
{(PAS) Max Planck Intitut f\"ur Gravitationsphysik
(Albert-Einstein-Institut), D-14476 Potsdam, Germany}
\altaffiltext{2}
{(MF) Institute of Astronomy, University of Cambridge, Madingley
Road, CB3~0HA Cambridge, UK}





\label{firstpage}

\begin{abstract}
\Rem{
Hubble Space Telescope (HST) observations reveal that young massive
star clusters form in gas-rich environments such as the Antenn{\ae}
galaxy. }
{Observations suggest that star clusters often form in binaries or
larger bound groups. Therefore, mergers between two clusters are likely to occur.}
If these clusters both harbor an
intermediate-mass black hole (IMBH; $\sim 10^{2-4}\,M_\odot$) in their
center, they can become a strong source of gravitational waves when
the black holes merge with each other.  In order to understand the
dynamical processes that operate in such a scenario, one has to study
the evolution of the merger of two such young massive star clusters,
and more specifically, their respective IMBHs.  We employ the
direct-summation {\NB} numerical tool on special-purpose GRAPE6
hardware to simulate a merger of two stellar clusters each containing
63,000 particles and a central IMBH. This allows us to study
accurately the orbital evolution of the colliding clusters and the
embedded massive black holes.  Within $\sim 7$~Myr the clusters have
merged and the IMBHs constitute a hard binary.  The final coalescence
happens in $\sim 10^8$ yrs. The implication of our analysis is that
intermediate-mass black holes merging as the result of coalescence of
young dense clusters could provide a source for the Laser
Interferometer Space Antenna ({\LISA}) space-based gravitational wave
detector mission. 
{We find that interactions with stars
increase the eccentricity of the IMBH binary to about 0.8. Although
the binary later circularizes by emission of gravitational waves, the
residual eccentricity can be detectable
through its influence on the phase of the waves if the last few years
of inspiral are observed.}
For proposed higher-frequency
space-based missions such as the Big Bang Observer ({\BBO}), whose first
purpose is to search for an inflation-generated gravitational waves
background in the $10^{-1}-1$ Hz range, binary IMBH inspirals would be
a foreground noise source.  However, we find that the inspiral
signals could be  characterized accurately enough that they could be
removed from the data stream and in the process provide us with
detailed information about these astrophysical events.

\end{abstract}

\keywords{Black hole physics, gravitational waves, stellar dynamics, methods:
N-body simulations}

\maketitle

\section{Introduction}

{There are at least two lines of evidence indicating that stellar
clusters, containing $10^4-10^6$ stars, can merge with each other relatively
early in there evolution.  First,} high-resolution Hubble Space Telescope
observations of the Antenn{\ae}  \citep{WhitmoreEtAl99,ZhangFall99} or
Stephan's Quintet \citep{GallagheretAl01} reveal hundreds of young massive star
clusters in the star forming regions.  These clusters are clustered into larger
complexes of a few 100 pc.  Since they harbor $\sim 10^5$ stars within a
few parsecs and are older than 5 Myr, they are most likely bound
clusters.  These {\em cluster complexes} have been suggested as the progenitors
of ultra-compact dwarf galaxies (UCDGs), as a result of the amalgamation of
tens or hundreds of their member clusters \citep{FK02a,FK05}.
{Furthermore, in more quiescent environments, a significant
fraction of clusters may form as bound binaries or low-order multiples.}
Observationally, this is indicated by the large number of young binary clusters
observed in the Magellanic Clouds (MCs). For instance, \citet{DMG02} estimate
that about one cluster in eight in the Large Magellanic Cloud is a member of a
bound group. Most clusters in (candidate) binaries are coeval and younger than
300\,Myr, suggesting that binary clusters generally merge early. Another
indication that mergers may be common, at least in environments such as the
MCs, is that the clusters in these galaxies are significantly flattened
\citep[e.g.,][]{KKSS89,KKSSD90,vandenBergh91}, possibly as a result of rotation
that could stem from a merger. The typical lack of significant
rotation in Galactic or M31 globular clusters \citep{HR94} does not invalidate
the idea that a large fraction of them may also be merger products; the more
intense tidal field of the Galaxy may lead to faster decrease of the rotation.
Theoretically, binary clusters are predicted to be a common outcome of the
off-center collision of two molecular clouds \citep{FK97,BBFC04} or the
collapse of a spherical shell of stars whose formation is triggered by a SN
explosion in a molecular cloud \citep{Theis02}.

{In this paper, we consider the possibility that the merging
clusters each contain an {\em intermediate-mass black hole} (IMBH; mass $M\sim
10^{2-4}\,M_\odot$) and investigate the consequences for gravitational-wave
(GW) astronomy. The observational evidence for IMBHs in clusters is only
suggestive \citep{vdM04}. However several authors have shown how mass
segregation leads in sufficiently compact young clusters to a phase of runaway
collisions among the heaviest stars, possibly leading to the formation of an
IMBH \citep{PortegiesZwartMcMillan00,GurkanEtAl04,PortegiesZwartEtAl04,
FreitagEtAl06}.}

\section{Formation of a merged cluster}

To form a realistic merged stellar cluster we make two clusters collide
on a parabolic orbit so that the minimum distance at which they pass by is
$d_{\rm min}$ if they are considered to be point particles at their respective
centers of mass.  We chose for our fiducial simulation $d_{\rm min}=2$ pc,
corresponding to a relative velocity at pericenter of $23.3 \kms$.  Given that
the clusters are set initially on a parabolic orbit, the inevitable loss of
energy (to, e.g., escaping stars) means that they will always merge to form a
larger cluster. Each cluster contains $\Npart = 6.3\times
10^{4}$ particles of $1\,\msol$ each which are
distributed according to King models with $W_0=7$ (cluster 1) and
$W_0=6$ (cluster 2). Their core radii are $R_{\rm core\,1}=0.203$ pc and
$R_{\rm core\,2}=0.293$ pc. Thus, they are compact enough to have experienced an early
core collapse that may lead to IMBH formation according to
\cite{GurkanEtAl04} and \cite{FreitagEtAl06} who find that for clusters with a
moderate initial concentration ($W_0\sim 6-7$), a half-mass radius
smaller than $1-2$\,pc is required. We note that our results, to be
presented ahead, can be rescaled to more extended clusters. The
central velocity dispersions are of $\sigma_{\rm core\,1}=8.41 \kms$
and $\sigma_{\rm core\,2}=8.29 \kms$. Both clusters harbor an
intermediate-mass black hole (IMBH) with a mass of $300
\msol$, so that the total particle number is $\Npart = 126\,002$. The
calculations were performed on special-purpose GRAPE-6A PCI
cards with the direct-summation $\NB$ code of \cite{Aarseth03}.  The particular
advantage of this code for our study is Aarseth's close-encounter
regularization scheme, which is free of any softening.  The peak performance of
these cards is of 130 Gflop, roughly
equivalent to 100 single PCs, which makes possible simulations of stellar
clusters with a realistic particle number.

%

\section{Evolution of the IMBH binary}


\begin{figure}
\resizebox{\hsize}{!}{\includegraphics[scale=1,clip]
{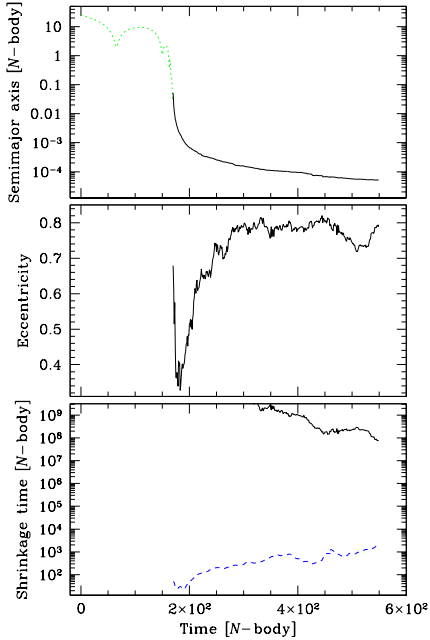}}
\caption{Evolution of the IMBH binary of the $N-$body simulation. 
In the upper panel, we show the shrinkage of the semi-major axis as a
solid line. The initial part of the curve in dashes represents the
separation between the IMBHs before they have formed a bound
binary. In the middle panel we display the evolution of the
eccentricity, which increases to $\sim 0.8$. The bottom panel gives a
an indication of the relevance of GW emission in the evolution of the
binary in comparison with stellar dynamical processes: the solid line
corresponds to the timescale for GWs emission \citep{Peters64} and the
dashed curve indicates the dynamical shrinkage timescale, given by
$a/\dot{a}$, where $\dot{a}$ is the time derivative of $a$. One sees
that at the end of the simulation when $a\simeq 5\times 10^{-5}\,\pc$,
GW emission is still less efficient than interaction with stars by
several orders of magnitude. 
\label{fig.EvolBin}
}
\end{figure}

\begin{figure}
\resizebox{\hsize}{!}{\includegraphics[scale=1,clip]
{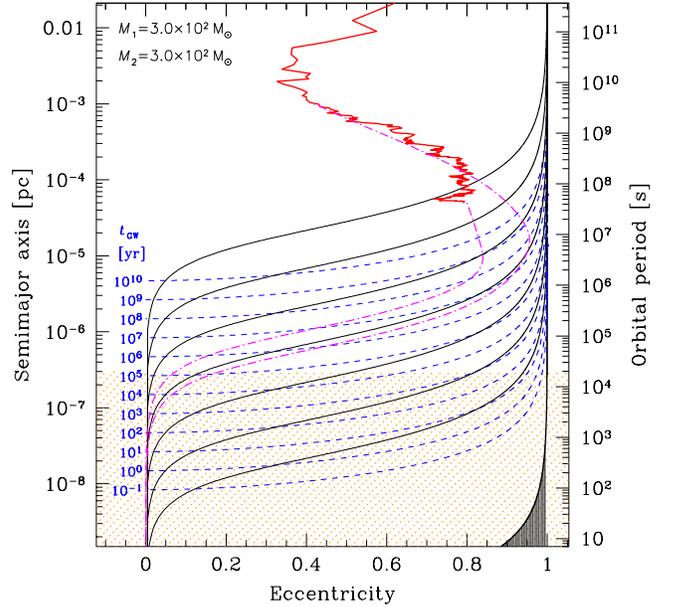}}
\caption{
Inspiral of an IMBH binary followed in the eccentricity--semi-major
axis plane. The jagged solid line (in red in the color online version)
is the result of the $N-$body simulation. The
dash-dotted lines (in magenta) are results of the analytical model
(equations \ref{eq.split}-\ref{eq.dedt}) for two sets of
parameters. In one case, we use $K_0=0.5$ and start at an early stage,
$a=10^{-3}\,$pc, $e=0.42$. The other track is for $K_0=0.1$ and starts
with the final values of the $N-$body run, $a=5\times 10^{-5}\,$pc,
$e=0.8$. In both cases the parameters for the stellar background --estimated from the $N-$body run-- are
set to $\sigma=23\,\kms$ and $\rho=1.6\times 10^5\,\Msun{\rm
pc}^{-3}$. The smooth solid (black) lines are trajectories for
evolution by GW emission \citep{Peters64} and the dashed (blue) lines
indicate the corresponding timescale, $t_{\rm GW}$ (labels on the
left). The dark dashed area, at the bottom right, indicates
approximately the region of unstable orbits. Finally the light (orange)
shaded area shows when the $n=2$ harmonic of the GW signal is in the LISA bandwidth
($P_{\rm orb}<2\times 10^4$\,s).
\label{fig.BinEvolWo17126k}
}
\end{figure}

In order to have a global view of the evolution of the system, in
Fig.(\ref{fig.EvolBin}) we present the dynamical evolution of the
{IMBH binary (hereafter ``binary'')}.  {The length and
time units used are $R_{\rm NB}=1\,$pc and $t_{\rm NB}=4.16\times 10^4\,$yr.}
In Fig.(\ref{fig.BinEvolWo17126k}) we show both the semi-major axis of the
binary and its orbital period versus its eccentricity.  Whilst the direct
summation simulations provide us with a highly accurate description of the
orbital parameters of the binary as it approaches the {\LISA} band, the
required computational effort to follow the evolution up to the very final
stage of coalescence is unjustified. Therefore in
Fig.(\ref{fig.BinEvolWo17126k}) we stop the direct-summation calculation after
the initial strongly fluctuating phase; when the eccentricity 
{achieves a large and seemingly steady value of $\sim 0.8$. We note that an
eccentricity larger than 0.65 is reached in additional simulations we have
performed with slightly different initial conditions and that a similar result
($e\simeq 0.8$) was found in a simulation of a merger of two clusters of
10,000 particles each using a special regularization method for the massive
binary (S.~Aarseth, personal communication; see \citealt{MA02} for the
numerical method).} In order to determine the binary properties in the {\LISA}
frequency band, we employ the results of the $N-$body simulation at
$t\simeq 550\,t_{\rm NB}\simeq 23$\,Myr and extend it with a analytical method
down to the moment in which the binary enters {\LISA} band. The basic idea
is to split the evolution of the semi-major axis and the eccentricity into
two contributions, one driven by the dynamical interactions with stars
(subscript {dyn}) and the other due to emission of GWs (subscript {GW}),

\begin{equation}
\frac{da}{dt}=\frac{da}{dt}\Big|_{\rm dyn}+\frac{da}{dt}\Big|_{\rm GW},~
\frac{de}{dt}=\frac{de}{dt}\Big|_{\rm dyn}+\frac{de}{dt}\Big|_{\rm GW}
\label{eq.split}
\end{equation}

\noindent
The ``GW'' terms are as given in \cite{Peters64}.  Using the 
relationships of
\cite{Quinlan96}, we have that since ${d(1/a)}/{dt}|_{\rm
dyn}=-{a^{-2}}\,{da}/{dt}= -H\,{G\rho}/{\sigma}$,

\begin{equation}
~\frac{da}{dt}\Big|_{\rm dyn}=-H\frac{G\rho}{\sigma}a^2.
\end{equation}

\noindent
In this last expression, $G$ is the gravitational constant, $\rho$ is
the density, $\sigma$ is the velocity dispersion and $H$ is the {\em
hardening constant} \citep{Quinlan96}. We measure $H\simeq 16$, in
agreement with previous works \citep{Quinlan96,SesanaEtAl04}.
 For a hard enough binary,
${de}/{d\ln(1/a)}|_{\rm dyn}=K(e)$ and so 
\begin{equation}
\frac{de}{dt}\Big|_{\rm dyn}=H\frac{G\rho}{\sigma}a\,K(e),
\label{eq.dedt}
\end{equation}

\noindent
with $K(e)\sim K_0\,e(1-e^2)$ (\citealt{MM05} and references
therein). We choose the value of $K_0=0.1$ for matching purposes with
the direct-summation $N-$body simulation. During the run, the ratio
$\rho/\sigma$ decreases by less than 30\,\% at the radius of influence
of the binary so we assume it is constant.  The evolution is dominated
by GW emission during the last $10^8$ yrs (the evolution from the
moment the binary is bound takes $\sim 160$ Myr) and enters the
{\LISA} bandwidth with $e\simeq 0.07$.  If we employ the fiducial
value of $K_0=0.5$ \citep{MM05} but start earlier on the same curve
(at a lower eccentricity), we find a evolution which fits the initial
part of the $N-$body run satisfactorily but produces too large an
eccentricity at later times.
\Rem{
we can see that although initially
it matches well the $N-$body simulation, it does not lead to the
same final eccentricity when entering the band, even if the
value is also low.  For the other simulations performed we do
not find any substantial difference from our fiducial case.
In the next section we discuss whether such eccentricities
are detectable.
}
\section{Detection by  {\LISA} and the {\BBO}}

\begin{figure}
\resizebox{\hsize}{!}{\includegraphics[scale=1,bb=18 150 570 690,clip]
{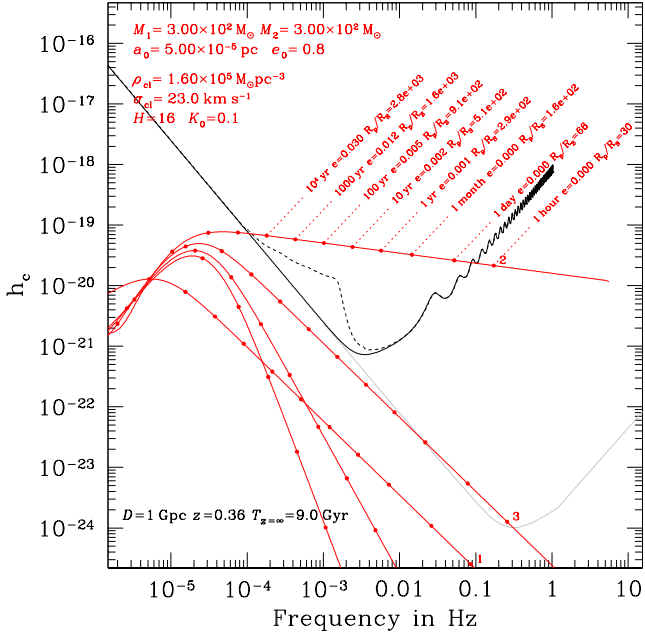}}
\caption{
\label{fig.amplitude}
Characteristic amplitude of the first five harmonics of the quadrupolar
gravitational radiation emitted during the inspiral of an
$300\,\Msun+300\,\Msun$ binary. The orbit is evolved according to
Eq.~\ref{eq.split}, starting with the conditions similar to those
reached at the end of the fiducial simulation, i.e., $e=0.8$,
$a=\tento{5}{-5}\,$pc with an ambient stellar density $\rho_{\rm cl}$ and (1-D)
velocity dispersion $\sigma_{\rm cl}$. We assume the source is at a distance
$D=1\,$Gpc. We indicate the noise curve $\sqrt{f\,S_h(f)}$ for {\LISA}
\citep{LHH00,Larson03}, with the Galactic binary white dwarf confusion
background in dashed line \citep{BH97}, and (in gray) for the {\BBO}
sensitivity curve \citep{CH06}. We label the position of the source at
various times before plunge. Note that the height of the point for the
amplitude above the {\LISA}/{\BBO} curve does {\em not} represent the SNR
(see text).}
\end{figure}

In the late phase of its inspiral, a binary may become a
detectable source of GWs. The characteristic amplitude of the
gravitational radiation from a source emitting at frequency $f$ is
$h_{\rm c} = (\pi D)^{-1}(2\dot E/\dot f)^{1/2}$, where $D$ is the
distance to the source, $\dot E$ is the power emitted and 
$\dot f$ the time derivative of the frequency \citep{FT00}.
\Rem{Strictly this corresponds not to the strain amplitude, but to the
strain amplitude {\em scaled} with the amount of time the source
spends in a given frequency bin; i.e.  $\sim f/\dot f$ (if $\sim
f/\dot f \ll$ duration of the mission). }With this definition, the
signal-to-noise ratio (SNR) of an event is obtained, assuming ideal
signal processing, by the integral $({\rm
SNR})^2=\int_{f_1}^{f_2}[{h^2_{\rm c}(f)}/{f\,S_h(f)}]\,d(\ln f)$,
where $f_1$ and $f_2$ are the initial and final frequencies of the
source during the observation and $S_h(f)$ is the instrumental noise
of the detector at frequency $f$ \citep{Phinney02,BC04}.

In Fig.~\ref{fig.amplitude} we follow the signal emitted by the IMBHs
of the fiducial simulation during their GW-driven inspiral. We plot
the five lowest harmonics of the quadrupolar emission \citep{PM63}. In
this figure, we assume a distance of 1\,Gpc. Both for {\LISA} and the
{\BBO}, only the $n=2$ harmonic is detectable, during the last few years
of inspiral. {However, the small residual excentricity
induces a difference in the phase evolution of the $n=2$ signal
compared to a circular inspiral. If the source is followed from a time
$t_{\rm mrg}$ before merger until merger, the accumlated phase shift
is 
\begin{equation}
\Delta \psi_e \simeq 1.0\left(\frac{e_{10^{-4}\rm Hz}}{0.05}\right)^2
\left(\frac{t_{\rm mrg}}{1\,\rm yr}\right)^{17/12}
\left(\frac{{\cal M}_z}{1000\,\Msun}\right)^{25/36},
\end{equation}
where $e_{10^{-4}\rm Hz}$ is the eccentricity when the $n=2$ signal
has reached a frequency of $10^{-4}$\,Hz and ${\cal M}_z \equiv
(1+z)(M_1M_2)^{3/5}(M_1+M_2)^{-1/5}$ is the redshifted chirp mass
\citep{CH06}. For our fiducial case (with $e_{10^{-4}\rm Hz}=0.07$), 
the eccentricity should be detectable ($\Delta \psi_e\ge 2 \pi$) if
observations span at least the last 3-4 years before merger.}
\Rem{The source needs to be closer than $\sim 50-100\,$Mpc
for the eccentricity to be detectable with {\LISA}. Even making
optimistic assumptions about the rates of binary IMBH formation (see
below), this distance is too close to encompass any source over a few
years of operation.  Considering more massive IMBHs does not make
eccentricity detection much easier despite the higher $h_{\rm c}$,
because of the decrease in frequency. Only with the {\BBO} and IMBHs
more massive than about $1000\,\Msun$ can one hope to measure a
non-zero eccentricity at 1\,Gpc.}A measured non-zero eccentricity is
an important constraint on the formation process and stellar
environment of the binary. The {\BBO} could detect a $300\,\msol +
300\,\msol$ inspiral to at least a redshift of 17 but the residual
eccentricity will not be measurable ($t_{\rm mrg}\le 1\,$day) and
significantly more massive IMBHs cannot be seen because their GWs are
redshifted to too low frequencies.

We now proceed to estimate roughly the detection rate of binary
IMBHs. \citet{FregeauEtAl06} have recently considered the
detection of binary IMBHs by {\LISA}, assuming that, generically, two
IMBHs form in any cluster undergoing a collisional runaway, as
suggested by the stellar dynamical simulations of
\citet{GFR06}. Here we assume that the runaway process leads to the
formation of only one IMBH; indeed it is not clear whether the two
collision-grown very massive stars (VMSs, with $\Mstar\gg 100\,\Msun$) can avoid merging with each other
before they become IMBHs. Compared to the scenario
considered by these authors, ours involve one more step, namely, the
merger of a IMBH-hosting cluster with another one. Schematically, the
computation of the detection rate for both scenarios can be written
\begin{equation}
\Gamma =  \frac{4\pi}{3}D_{\rm max}^3 \times \dot{n}_{\rm clust} \times P_{\rm bin}.
\label{eq.rate}
\end{equation}

\noindent $D_{\rm max}$ is the distance at which a
coalescence event can be detected, $\dot{n}_{\rm clust}$ is the rate
of formation per unit volume of clusters massive enough
for (potential) IMBH formation. Rough estimates for these terms are
$D_{\rm max}\approx 1-3\,$Gpc and $\dot{n}_{\rm clust}\gtrsim 5\times
10^{-10}\,{\rm Mpc}^{-3}{\rm yr}^{-1}$ \citep{FregeauEtAl06}. $P_{\rm
bin}$ is the probability that a cluster will host a (coalescing) IMBH
binary. Only this term differs between the two scenarios. In our
case, it can be written as 
\[
P_{\rm bin}=P_{\rm merg} P_{\rm ra}^2 P_{\rm IMBH}^2
\]
where $P_{\rm ra}$ is the probability that a cluster
evolves to the runaway phase, $P_{\rm IMBH}$ is the probability that
the runaway leads to IMBH formation and $P_{\rm merg}$ is the
probability that the cluster merges with another cluster. For
\citet{FregeauEtAl06}, it is simply $P_{\rm bin}=P_{\rm ra}
P_{\rm IMBH}$. Eq.~\ref{eq.rate} is an oversimplification because the
various factors are not independent of each other; for instance,
$D_{\rm max}$ is larger for more massive IMBHs, which probably introduces
a dependence on $M_{\rm cl}$. Also the cosmological evolution of the
cluster formation rate should be taken into
account. \citet{FregeauEtAl06} present a much more rigorous
computation but we can use the above equations to derive our rate
estimate from their results, $\Gamma 
\approx P_{\rm merg}P_{\rm ra}\,\Gamma_{\rm Fregeau}$ with $\Gamma_{\rm Fregeau}\approx 40-50(P_{\rm ra}/0.1)\,{\rm yr}^{-1}$.

We take the same optimistic view as Fregeau et al.\ that VMSs always
form IMBHs, $P_{\rm IMBH}=1$. The extra $P_{\rm ra}$ factor indicates
that both clusters must contain an IMBH. The conditions for
runaway are relatively well understood
\citep{GurkanEtAl04,PortegiesZwartEtAl04,FreitagEtAl06,GFR06} but the
value of $P_{\rm ra}$ is highly uncertain because these conditions
apply to the {\em initial} cluster properties.  Fregeau et
al.\ choose $P_{\rm ra}=0.1$ as an illustrative value. We note that
if all clusters are born with a concentration as high as a $W_0=8$
King model and the mass and half-mass radii of observed
local globular clusters are similar to their initial values, $P_{\rm
ra}$ may be about 0.5 \citep{FreitagRB06}. On the other hand,
\citet{BME04b} have argued that Galactic globular clusters containing
an IMBH may be amongst the least dense clusters (for their mass) as a
result of the strong IMBH-powered gravothermal expansion (see also 
\citealt{FASK06}). The
relatively small number of observed low-density clusters would
therefore suggest a smaller $P_{\rm ra}$ but the galactic tidal field
may prevent the expansion of the cluster.

The factor $P_{\rm merg}$ is also uncertain. As mentioned in the
introduction, observational and theoretical points support the
possibility of a large fraction of clusters being born in binaries,
suggesting $P_{\rm merg}=0.1-1$.  Furthermore, if the scenario of
Fellhauer \& Kroupa holds, each UCDG would require $\sim 10-100$
cluster-cluster mergers. UCDGs may be as numerous as normal galaxies,
at least in galaxy clusters
\citep[e.g.,][]{HM04,DrinkwaterEtAl05}. Hence the number of clusters
that have been incorporated into UCDGs may be of the same order of
magnitude as the number of those that have survived as isolated
clusters (possibly after a binary merger). This indicates that the
rate estimate for the ``UCDG channel'' should be similar (given the
considerable uncertainties) to that of the ``binary-merger channel''.
Finally, we find that the detection rate estimate for {\LISA} is of
\[
\Gamma=4-5(P_{\rm ra}/0.1)^2 P_{\rm merg}\,{\rm yr}^{-1}.
\]

Because of its superior sensibility at frequencies higher than $\sim
10^{-3}\,$Hz, the {\BBO} should be sensitive to IMBH binary coalescences
from redshifts to at least $\sim 15$. A high merger rate could be
envisaged as a potential confusion foreground for this
mission. Fortunately, following similar arguments as discussed in
\cite{CH06} regarding the contribution of neutron star mergers, 
IMBH binaries should be easy to subtract out (C.~Cutler personal
communication). Thus the {\BBO} will open a new window of possibilities
for our understanding of astrophysical scenarios leading to
coalescence of IMBH binaries.

\section*{Acknowledgments}

We thank Leor Barack, Curt Cutler and Jonathan Gair for enlightening
discussions and Cole Miller for helping us improving the manuscript.  We
acknowledge the Astronomisches Rechen-Institut for the computing resources on
the GRACE cluster of the Volkswagen Foundation, SFB439. The work of PAS has
been supported in the framework of the Third Level Agreement between the DFG
and the IAC (Instituto de Astrof\'\i sica de Canarias). The work of MF is
funded through the PPARC rolling grant at the Institute of Astronomy (IoA) in
Cambridge. PAS is indebted with the IoA for inviting him the month of February
2006, where MF and he started working on this project. 


\label{lastpage}

\end{document}